\documentclass[10pt]{iopart}
\usepackage{graphicx}
\usepackage{amsmath, amssymb}

\newcommand \be{\begin{eqnarray}}
\newcommand \ee{\end{eqnarray}}
\newcommand \ba{\begin{eqnarray}}
\newcommand \ea{\end{eqnarray}}
\renewcommand \bs{\boldsymbol}
\newcommand \mc{\mathcal}
\newcommand \mE{\mathcal E}
\newcommand \E{{\bs{\mc E}}}

\def\nn{\nonumber}

\begin{document}
\title[Diffusion correction to the Raether-Meek criterion]
      {Diffusion correction to the Raether-Meek criterion for
      the avalanche--to--streamer transition} 
\author{Carolynne Montijn$^1$, Ute Ebert$^{1,2}$}
\address{$^1$CWI, P.O.Box 94079, 1090 GB Amsterdam, The Netherlands,}
\address{$^2$Dept.\ Physics, Eindhoven Univ.\ Techn., The Netherlands.}

\date{\today}

\begin{abstract}
Space-charge dominated streamer discharges can emerge in free space from
single electrons. We reinvestigate the Raether-Meek criterion
and show that streamer emergence depends not only on ionization 
and attachment rates and gap length, but also on electron diffusion. 
Motivated by simulation results, 
we derive an explicit quantitative criterion for the avalanche-to-streamer 
transition both for pure non-attaching gases and for air, 
under the assumption that the avalanche emerges from a single free 
electron and evolves in a homogenous field.
\end{abstract}

\maketitle

\section{Introduction}

\subsection{Problem setting and review}

Emergence and propagation of streamer-like discharges are topics
of current interest. Streamers play a role in creating the 
paths of sparks and lightning~\cite{maz1995,baz2000}
and in sprite discharges at high altitude above 
thunderclouds~\cite{ger2000,pas2002,liu2004}.
They are also used in various industrial applications~\cite{che1995}, 
e.g., in corona reactors for water and gas 
treatment~\cite{eli1991,shi1997,lis1999,win2005},
and as sources of excimer radiation for material 
processing~\cite{mak1998,oda2000,kog2002}, for a recent
overview see~\cite{ebe2006}.

In the present paper, we investigate the conditions 
under which a tiny ionization seed as, in particular, 
a single electron in a homogeneous electric field
far from any electrodes grows out into a streamer with self-induced 
space charge effects and consecutive rapid growth. 
The critical length of time for this transition as a function of 
the electric field is usually described by the Raether-Meek
criterion. We will confront current simulation results
with the underlying assumptions of the Raether-Meek criterion,
and then derive a diffusion correction to it. This correction can
amount to a factor of 2 or more for transition time and length 
for certain parameters as we will elaborate below and summarize
in Figs.~\ref{transED} and \ref{ftair}. 
The consequences are particularly
severe in non-attaching gases, where in low fields the diffusion
can suppress streamer formation almost completely while the Raether-Meek
criterion would predict streamer formation after a finite travel
distance and time. An example of such an avalanche in extremely low fields 
is discussed in~\cite{dow2003}.

In many applications, discharges are enclosed by containers and
electrodes; streamers then frequently emerge from point or rod
electrodes, that create strong local fields in their 
neighborhood~\cite{phe1976} and also influence the discharge by
surface effects. 
On the other hand, in many natural discharges and, in particular, 
for sprites above thunderclouds~\cite{liu2004}, it is appropriate 
to assume that the electric field is homogeneous and metal electrodes 
absent. In this case, single electrons can create ionization avalanches
that move into the electron drift direction. From those avalanches,
single or double ended streamers can emerge, and we are interested in the
prediction of this transition. For clarity, we call a spatial
distribution of charged particles an avalanche, if the electric field
generated by their space charges is negligible in comparison to
the background external field; on the other hand, if the space charges 
of the distribution substantially contribute to the total field, 
we speak of a streamer.

The critical field required for lightning generation is presently
a topic of debate, in particular, whether thundercloud fields
are sufficient for classical breakdown or whether relativistic
particles from cosmic air showers are required~\cite{dwy2003,gur2005}.
Different critical fields can be defined for different processes; 
e.g., in \cite{phe1976} a critical field for the propagation 
of positive streamer propagation is suggested that is valid 
after the streamers have emerged from a needle electrode.
This field is certainly lower than the critical field for
streamer emergence from an avalanche to be discussed here.

Of course, dust particles or other nucleation centers can play 
an additional role in discharge generation in thunderclouds, 
but in the present paper we will focus 
on the effect of a homogeneous field in a homogeneous gas. 
This assumption corresponds to the classical experiments of Raether
in the thirties of the last century~\cite{rae1939}. 

Within the present introductory and motivating section, we first recall 
the common discharge model 
and present simulation results for avalanches and consecutive streamers
that emerge from a single electron in a homogeneous field far from 
any surfaces. Then we recall the Raether-Meek 
criterion; it suggests that the avalanche to streamer 
transition depends on the ionization rate $\alpha$ and gap length $d$ 
through the dimensionless combination $\alpha d$.
We confront this criterion with our simulations and argue that 
the transition depends not only on the ionization coefficient 
times gap length
but also on electron diffusion. Now numerical evaluations
of the initial value problem for a large range of parameters,
namely fields, gas types and densities,
would be very tedious. However, we have succeeded in making
analytical progress on the transition criterion.
This has two major advantages: first, general expressions for arbitrary
fields, gases and densities can be derived. Second, the result
can be given in the form of a closed mathematical expression.
These calculations and results form the body of the paper.

\subsection{Discharge model and simulation results}

In detail, we consider a continuous discharge model with attachment and 
local field-dependent impact ionization rate and space charge effects.
It is defined through 
\ba
\partial_t\,n_e&=& 
 \nabla_{\bf R}\cdot(D_e\nabla_{\bf R}n_e+\mu_e\,{\bf E}\, n_e) 
+(\mu_e\,|{{\bf E}}|\,\alpha_i(|{\bf E}|)-\nu_a)\,n_e ,\label{ned}\\[2mm]
\partial_t\,n_+&=& 
\mu_e\,|{{\bf E}}|\,\alpha_i(|{\bf E}|)\,n_e ,\label{n+d}\\[2mm]
\partial_t\,n_-&=&\displaystyle\nu_a\, n_e ,\label{n-d}\\[2mm]
\nabla_{\bf R}^2\Phi&=&\displaystyle\frac{\rm e}{\epsilon_0}\;(n_e+n_--n_+)~~~,~~~
{{\bf E}}=-\nabla_{\bf R}\Phi,\label{phid}
\ea
where charged particles are present only in a bounded region,
and the electric field far away from the ionized region is homogenous.
Here $n_e$, $n_+$ and $n_-$ are the particle densities of electrons,
positive and negative ions, and 
${{\bf E}}$ and $\Phi$ are the electric field and potential, 
respectively. The total field ${\bf E}$ is the sum of the background 
(Laplacian) field ${\bf E}_b$ in the absence of space charges 
and the field generated by the charged particles ${\bf E}'$. 
$D_e$ and $\nu_a$ are the electron diffusion 
and the electron attachment rate, respectively. 
The impact ionization coefficient $\alpha_i$ is a function of the electric field, 
as established in various books, and for our numerical calculations, 
we use the Townsend approximation 
\be
\alpha_i(|{{\bf E}}|)=\alpha_0\exp(-E_0/{|{\bf E}}|),
\label{alpha}
\ee
in which $\alpha_0$ and $E_0$ are parameters for the effective cross section.
They depend on the ratio of background and normal gas density 
($N$ and $N_0$, respectively) 
as $\alpha_0\propto (N/N_0)$ and $E_0\propto (N/N_0)$~\cite{rai1991}. 
This scaling is equivalent to stating that the reduced electric
field ${\bf E}/N$ is the relevant physical variable for impact
ionization processes.
The positive and negative ions are considered to be immobile on the time
scales investigated in this paper because avalanches and streamers 
evolve on the time scale of the electrons that are much more mobile
due to their much lower mass. 

\begin{figure}
\begin{center}
\includegraphics{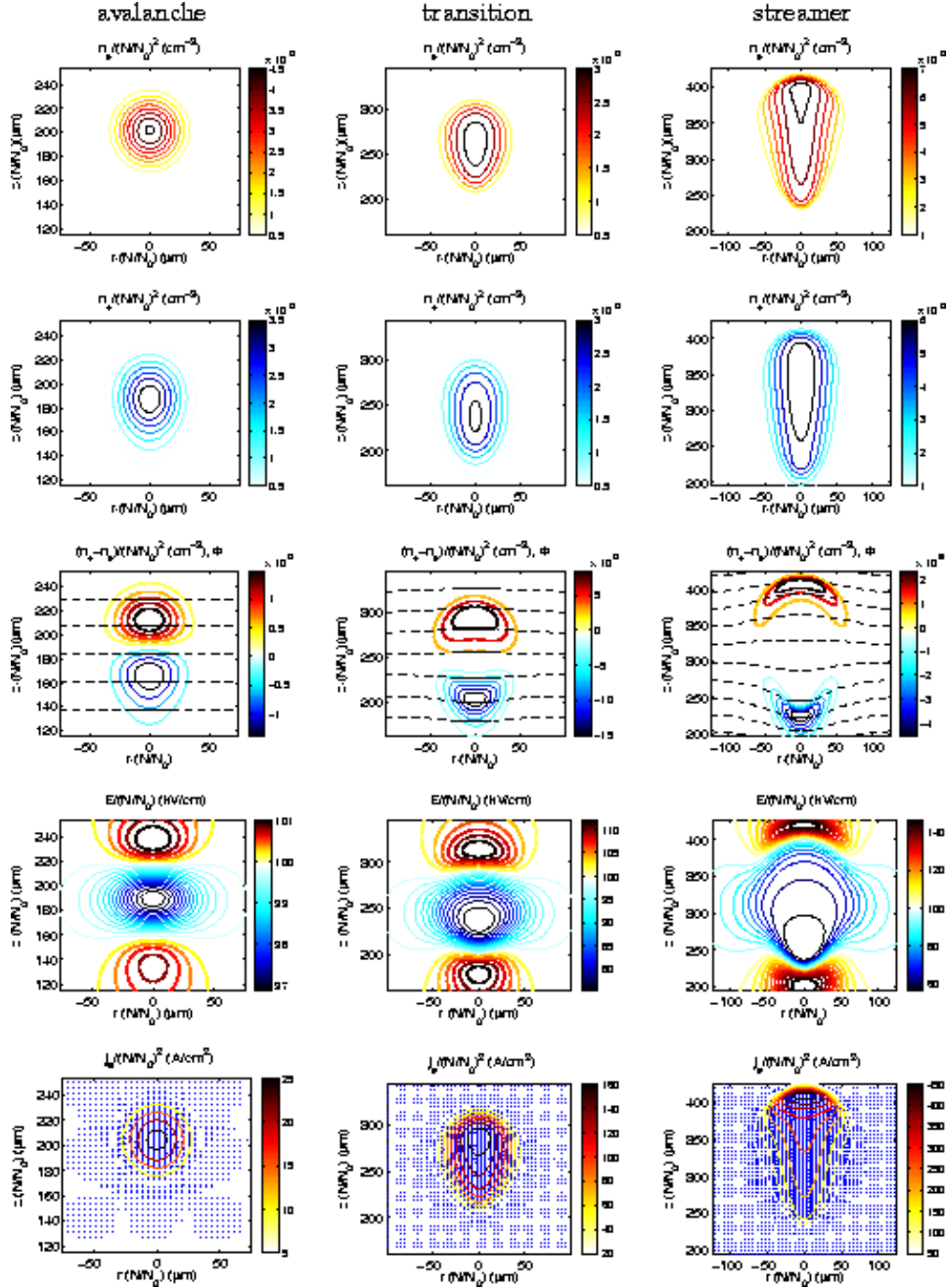} 
\caption{The avalanche to streamer transition:
numerical solution of the discharge model~(\ref{ned})-(\ref{phid})
for pure N$_2$ in a uniform background field, under
assumption of cylindrical symmetry. As N$_2$ is a non-attaching
gas, there is no formation of negative ions.
The background electric field is directed in the negative 
$z$-direction and has a strength $E_b/(N/N_0)=100$~kV/cm
where $N$ is the actual particle density and $N_0$ the 
particle density under normal conditions. 
Initially, a single electron is placed at a distance $z$=115 $\mu$m of
the cathode (which in turn is placed at $z=0$). 

Shown are the electron avalanche phase 
(left column), the transition to streamer (middle column) and 
the space charge dominated streamer phase (right column).
The respective times are $t$= 0.225, 0.375 and 
0.525 ps for N$_2$ under normal conditions.

From top to bottom: electron number density distribution; 
ion number density distribution; net charge density distribution 
(positive: thin lines, negative: thick lines) and equipotential
lines (dashed); electric field strength (smaller than 
the background field: thin lines, larger: thick lines); 
current density ${\bf j}_e=\mu_e{\bf E}n_e+D_e\nabla n_e$.
}
\label{plotav}
\end{center}
\end{figure}

We consider the situation where a tiny ionization seed of the size
of one or a few free electrons is placed in free space, i.e., 
within a gas far from walls, electrodes or other boundaries. 
If the externally applied field is sufficiently
high, it will develop into an electron avalanche that will drift 
towards the anode.
Eventually, the charged particle density in the avalanche will
become so large that space charge effects set in and change the
externally applied field. As a consequence, the interior of 
the formed very weak plasma will then be weakly screened from the external
field while the field at the outer edges is enhanced. Depending
on photo-ionization processes, then an anode-directed or a double 
ended streamer emerges from the avalanche.  
This evolution from an electron avalanche to a streamer is illustrated
in Fig.~\ref{plotav}. Details on our simulations can be found in
\cite{mon2005-2,mon2005-3,mon2006}, here we only use them
for illustration purposes.

Fig.~\ref{plotav} shows essential features of the solutions
that will be substantiated by quantitative analysis in the 
body of the paper.
In the left column, an avalanche can be seen:
the electron distribution (upper row) is Gaussian 
and spherically symmetric. The position of the Gaussian is 
determined by electron drift in the homogeneous background field,
its width by electron diffusion.
The ions (second row) are left behind (i.e., further down) and 
stretched along the temporal trace of the avalanche. The resulting 
space charge distribution (third row) is essentially a smooth
dipole without much structure. Actually, these pictures are quite similar
to the sketches of Raether. The electric field (fourth row)
is essentially unchanged up to corrections below 1\%.
The current (lowest row) shows the same Gaussian structure
as the electrons; it is dominated by electron drift 
$\mu_e n_e({\bf R},t)\;{\bf E}_b$ in the homogeneous background field
${\bf E}_b$ with a small diffusional correction.
In the right column, a conducting filament is formed, and the streamer stage 
is reached. Electron and ion distribution show a similar long stretched shape.
The space charges approach a layered structure, and the field ahead of 
the streamer is changed by these space charges by up to 40\%. 

There is some freedom in defining the transition point from avalanche
to streamer. In the body of the paper, we will argue that a maximal 
field enhancement of 3\% ahead of the streamer, i.e.,
\be
\label{k}
k=\frac{\max_{\bf R}|{\bf E}({\bf R},t)|-|{\bf E}_b|}{|{\bf E}_b|},~~~k=0.03
\ee
is a decent measure for the transition. We will see that essentially
up to this moment of time the total number of electrons in the avalanche
grows exponentially in time, while in the streamer phase,
the growth is slower. 

\subsection{Review of critical field and Raether-Meek criterion}

Essentially two criteria have been given in the literature for the
emergence of a streamer from a tiny ionization seed,
one for the required background field and one for the
required space and evolution time. The first one is 
a necessary lower bound for the background field: 
the electric field has to be higher than the threshold field 
$E_k$ where the impact ionization rate overcomes the attachment 
rate. The ionization level can only grow if the rightmost term in Eq.~(\ref{ned})
is positive, hence if the effective ionization coefficient is positive,
\be
\alpha(|{\bf E}|)=\alpha_i({\bf E}|)-\nu_a/(\mu_e|{\bf E}|)>0\, .
\label{alphaeff}
\ee
This determines the threshold field $E_k$ as
\be
\mu_e\;E_k\alpha_i(E_k)=\nu_a\, .
\label{Ek}
\ee

The second criterion is known as the Raether-Meek criterion. 
It states that the total electron number must have reached the order
of $10^8$ to $10^9$ for space charge effects to set in.
If this number is reached by exponential multiplication of one
initial electron within a constant field ${\bf E}_b$, this means that
\be
{\rm e}^{\alpha(|{\bf E}|)d}
\approx 10^8 \mbox{ to } 10^9 ,
\label{ad}
\ee
where $d$ is the avalanche length. In brief as a rule of thumb 
the criterion reads
\be
\alpha(|{{\bf E}}|)\; d \approx 18 ~{\rm to}~ 21
~~\mbox{ according to Raether and Meek}.
\label{meek}
\ee
Let us first note that the same criterion has been 
suggested for quite different situations in the literature. 
In his original article, Meek \cite{mee1940} studies the emergence of 
a cathode directed (i.e., positive) streamer from 
an anode directed avalanche that has bridged a short gap.
On the other hand, Bazelyan and Raizer~\cite{baz1998} study 
the emergence of streamers in free space, i.e. far away from 
the electrodes. 
To estimate the field of the ions, Meek used the diffusion radius 
of the electron avalanche, and the ionization rate
in the background field, however, the diffusion does not show
up in his transition criterion. Bazelyan and Raizer on the other hand, 
neglect diffusion and base their criterion on the radius 
of the avalanche due to electrostatic repulsion. 
All authors \cite{mee1940,baz1998,loe1940,loe1940-2} 
assume the electron distribution to be spherically
symmetric, on the other hand, they base their transition criterion
on a total field screening, i.e., to $k=1$ in Eq.~(\ref{k}).
In view of available simulation results like our Fig.~\ref{plotav}, 
these assumptions are self-contradictory.

Apart from these considerations on the history of the derivation,
there are actually two major reasons to revise the Raether-Meek-criterion:
\begin{itemize}
\item[1)] The prediction that a parameter should be in the range
of 18 to 21 (where authors seem to be willing to assume an
even larger range of values to get consistency with experiment) 
is not very satisfactory and invites improvement.
\item[2)] Diffusion has to be included into the model for physical
as well as for mathematical reasons. Without diffusion,
an initially concentrated electron package 
would not spread and it would create enormous fields within very short time
as they are well known in the neighborhood of point sources.
Indeed, diffusion decreases the electron density and the maximal fields
while impact ionization increases it. In low fields, diffusion stays 
dominant for a long time
and delays space charge effects and consecutive streamer emergence.
It is therefore clear that the avalanche to streamer transition
does not only depend on multiplication rates, but also on the relative
importance of diffusional spreading. This should provide a more quantitative
transition criterion than the pure Raether-Meek criterion.
\end{itemize}

\subsection{Organization of the paper}

We will derive a diffusion correction to the Raether-Meek criterion
through the following steps:
in Section 2, the intrinsic scales of the problem with their explicit
density dependence are identified through dimensional analysis. 
In Section 3, we analyze the spatial distribution of the electrons
during the avalanche phase and their contribution to the electric field; 
this gives a first approximate correction to the Raether-Meek-criterion.
In Section 4, we approximate the spatial distribution of the ions
and their contribution to the electric field. Electron and ion
field are then combined to give the total change of the electric field
during the avalanche phase. If this field becomes ``substantial''
(cf. Fig.~\ref{plotav} and Eq.~(\ref{k})), we have found the
avalanche-to-streamer transition.
Finally, the analytical non-dimensional results are translated 
back to dimensional quantities,
and we refer the reader interested in the final prediction only to 
Figs.~\ref{transED} and~\ref{ftair} for the transition criterion
in non-attaching gases and in air. These figures visualize the
analytical criterion (\ref{transt1}).  
Section 5 contains the conclusions, and an additional approximation
for the ion generated electric field can be found in the appendix.

\section{Dimensional analysis}

The Raether-Meek criterion can be understood as a simple example 
of dimensional analysis. Dimensional analysis identifies general physical
properties in terms of dimensionless numbers that are independent 
of a particular gas type or density. The physical importance 
of dimensionless numbers like the Reynolds number is well
known in hydrodynamics, we here follow the same approach.

In the light of dimensional analysis, the Raether-Meek criterion
states that the effective cross-section $\alpha(|{{\bf E}}|)$
has the dimension of inverse length, hence the dimensionless
number $\alpha(|{{\bf E}}|)d$ characterizes the gap length 
in multiples of the ionization length and therefore the exponential
multiplication rate $e^{\alpha d}$. This number directly characterizes 
the total number of electrons in an avalanche that started
from a single free electron. However, this is not the only 
dimensionless number in the problem, a second one is 
the dimensionless diffusion constant 
\be
\label{D}
D=\frac{D_e\alpha_0}{\mu_eE_0},
\ee
that plays a distinctive role in the avalanche to streamer 
transition as well as it determines the width of the electron cloud.
Note that this dimensionless diffusion constant is related to the 
electron temperature as $D_e/\mu_e=k_BT_e$ where $k_B$ is the
Boltzmann constant. The electron temperature $T_e$ actually 
can be defined through this relation,
even if the electron energy distribution is not Maxwellian 
in the presence of strong electric fields. Furthermore, $D$
depends on $\alpha_0/E_0$, the two parameters characterizing
the Townsend coefficient, given by Eq.~(\ref{alpha}),
 for the specific gas and density.

For the general performance of dimensional analysis, we refer
to earlier articles \cite{Ebe1997,roc2002} and here only state 
the results: lengths are measured in units of $1/\alpha_0$,
fields in units of $E_0$, velocities in units 
of $v_0=\mu_eE_0$ and time consistently in units of
$t_0=1/(\alpha_0\mu_eE_0)$ --- hence diffusion should 
be measured in units of $\mu_eE_0/\alpha_0$ as done in (\ref{D}).
The natural scale for the particle densities follows from the Poisson equation,
$n_0=\epsilon_0\alpha_0E_0/e$.

The parameters $\alpha_0$, $\mu_e$, $D_e$ and $E_0$ depend on the ratio
of the background gas density $N$ and the gas density under normal
conditions $N_0$. Using parameters as 
in~\cite{rai1991,dha1987,vit1994,dav1971},
the characteristic scales are for N$_2$:
\be
\begin{array}{ll}
\displaystyle \ell_0=\alpha_0^{-1}=2.3~{\rm \mu m}\;\frac{1}{N/N_0},&
\displaystyle~~~ E_0=200 ~\frac{\rm kV}{\rm cm}\;\frac{N}{N_0}, \\[4mm]
\displaystyle\mu_e= 380~\frac{\rm cm^2}{\rm V s}\;\frac{1}{N/N_0},&
\displaystyle~~~ D_e= 1800~\frac{\rm cm^2}{\rm s}\;\frac{1}{N/N_0},\\
\displaystyle v_0=7\cdot10^7~\frac{\rm cm}{\rm s},&
\displaystyle~~~t_0=3~{\rm ps}~\frac{N}{N_0},\\
\displaystyle n_0=4.8\cdot10^{14}~\frac{1}{\rm cm^3}
\left(\frac{N}{N_0}\right)^2,& ~\\
\end{array}
\label{coeff}
\ee
and the dimensionless diffusion constant is $D=0.1$.
Notice that the characteristic velocity scale is independent of pressure, 
in agreement with measurements of streamer velocities at different pressures.
Notice furthermore, that it directly follows from this analysis
that the relevant physical parameter is the reduced electric field 
${\bf E}/N$.

Dimensionless parameters and fields are introduced as 
\be
&&{\bf r}= \frac{\bf R}\ell,~~~ \tau=\frac t {t_0},
~~~\nu=\nu_a t_0,
\nonumber\\
&& \sigma=\frac{n_e}{n_0},~~~ \rho=\frac{n_+-n_-}{n_0},~~~
\bs{\cal E}=\frac{{\bf E}}{E_0},
\ee
which brings the system of equations (\ref{ned})-(\ref{phid}) into the dimensionless form
\ba
\partial_\tau\,\sigma & = &
   D\nabla^2\sigma+ \nabla(\bs{\cal E}\sigma)+ 
   f(|\bs{\cal E}|,\nu)\,\sigma \label{sigmand}\, ,\\
\partial_\tau\,\rho & = & f(|\bs{\cal E}|,\nu)\,\sigma \label{rhond}\, ,
\\
-\nabla^2\phi & = &\rho-\sigma\,~~~~~\bs{\cal E}=-\nabla\phi ,
\label{poissnd}
\ea
where the operator $\nabla$ is taken with respect to ${\bf r}$ and
where $f(|\bs{\cal E}|,\nu)$ is the dimensionless effective
ionization rate,
\be
f(|\bs{\cal E}|,\nu)=\frac{\mu_e\,|{\bf E}\,|\alpha(|{\bf E}|)}{\mu_e\, E_0\,\alpha_0}= |\E|e^{-1/|\E|}-\nu.
\ee
It is remarkable that the density of positive and negative ions 
$n_\pm$ enters the equations only in the form of the single dimensionless
field $\rho\propto n_+-n_-$. This is clear in the case of the Poisson
equation, but holds also for the generation term proportional
to $f(|\bs{\cal E}|,\nu)$. This coefficient
accounts for the production of free electrons through impact ionization 
and for the loss of free electrons due to attachment. 

We neglect the effect of photoionization as its rates are typically 
much lower than impact ionization rates; it does not contribute 
significantly to the build-up of a compact ionized cloud where
eventually space charge effects will set in (quite in contrast
to its distinct role in positive streamer propagation).

\section{Electron distribution and field}

We derive the transition as follows: We assume that an
avalanche starts from a single electron and follows a transition
as shown in Fig.~\ref{plotav}. In the calculation we
neglect space charge effects on the evolution of densities, 
but we do calculate the additional 
electric field generated by the space charges.
If this field reaches a relative value of k=0.03 --- this 
value will be motivated in Section 4.3 ---, space charge effects
are not negligible anymore, and the transition to the streamer
is found.

The electric field generated by space charges has one
contribution from the electrons $\sigma$ and another one
from the positive and negative ions $\rho$. In the present section,
we calculate the field of the electrons, in the next section,
we will include the field of the ions.

\subsection{The electron distribution: a Gaussian}

We write the single electron that generates the avalanche as a
localized initial density
\begin{equation}
\sigma({\bf r},\tau=0)=\rho({\bf r},\tau=0)=\sigma_0\delta({\bf r} -{\bf r_0})
\label{elinit}
\end{equation}
and consider its evolution under influence of a uniform field 
$\bs{\cal E}_{b}=-{\mc E}_b\hat{\bf e}_z$, where
$\hat{\bf e}_z$ is the unit vector in the $z$ direction and $\mE_b=|\E_b|$ is constant.
A single electron is written as a $\delta$-function
$n_e({\bf R})\propto\delta^3({\bf R}-{\bf R}_0)$ in physical units
where the spatial integral over the electron number density
\be
N_e(\tau)=\int d^3{\bf R}\;n_e({\bf R}),
\ee
of course, should be unity $N_e(0)=1$. According to the last section, this corresponds in
dimensionless units to $\sigma_0=1/(n_0\ell_0^3)$ which is 
$1.7\cdot 10^{-4}~N/N_0$ for nitrogen. We will use $\sigma_0=10^{-4}$
in the sequel. We emphasize, however, that the theory will be developed
for an arbitrary value of $\sigma_0$.

During the avalanche phase the electric field remains unaffected 
by space charges, so that the continuity equations for the charged
particles~(\ref{sigmand})-(\ref{rhond}) can be linearized around the
background field,
\ba
\partial_\tau\,\sigma & = &
   D\nabla^2\sigma+ \E_b\cdot\nabla\sigma+ 
   \sigma\,f \label{sigmandav}\, ,\\
\partial_\tau\,\rho & = & \sigma\,f \label{rhondav}\, ,
\ea
where $f=f(\mE_b,\nu)$.

For the initial condition~(\ref{elinit}), the electron evolution
according to Eq.~(\ref{sigmandav}) can be given explicitly as~\cite{rai1991}
\begin{equation}
\sigma({\bf r},\tau)=\sigma_0\,e^{f(\mE_b,\nu)\tau}\;
\frac{e^{-({\bf r}-{\bf r}_0 +\E_b\tau)^2/(4D\tau)}}{(4\pi D\tau)^{3/2}};
\label{sigmaav}
\end{equation}
it has the form of a Gaussian package that drifts with velocity 
$-\E_b$, widens diffusively with half width proportional to $\sqrt{4D\tau}$,
and carries a total number of electrons $\sigma_0e^{f(\mE_b,\nu)\tau}$. 
(If the initial ionization seed consists of several electrons in
some close neighborhood, the Gaussian shape is approached
nevertheless for large times due to the central limit theorem.)

Integrating Eq.~(\ref{sigmaav}) over the entire space shows
that the total number of electrons grows as 
$N_e(\tau)=\sigma_0n_0\ell_0^3e^{f\tau}$ 
(if we start with a single electron). On the other hand,
the maximum of the 
electron density is reached at the center of the Gaussian
${\bf r}={\bf r}_0 -\E_b\tau$ and has the value 
\be
\sigma_{\text{max}}(\tau)=\max_{{\bf r}}\sigma({\bf r},\tau)
=\frac{\sigma_0\,e^{f\tau}}{(4\pi D\tau)^{3/2}},
\label{sigmamaxth}
\ee
hence it first decreases until $\tau=3/(2f)$ due to diffusion 
and then increases due to electron multiplication.  
At this moment of evolution, generation overcomes diffusion.

\begin{figure}
\begin{center}
\includegraphics[width=12cm]{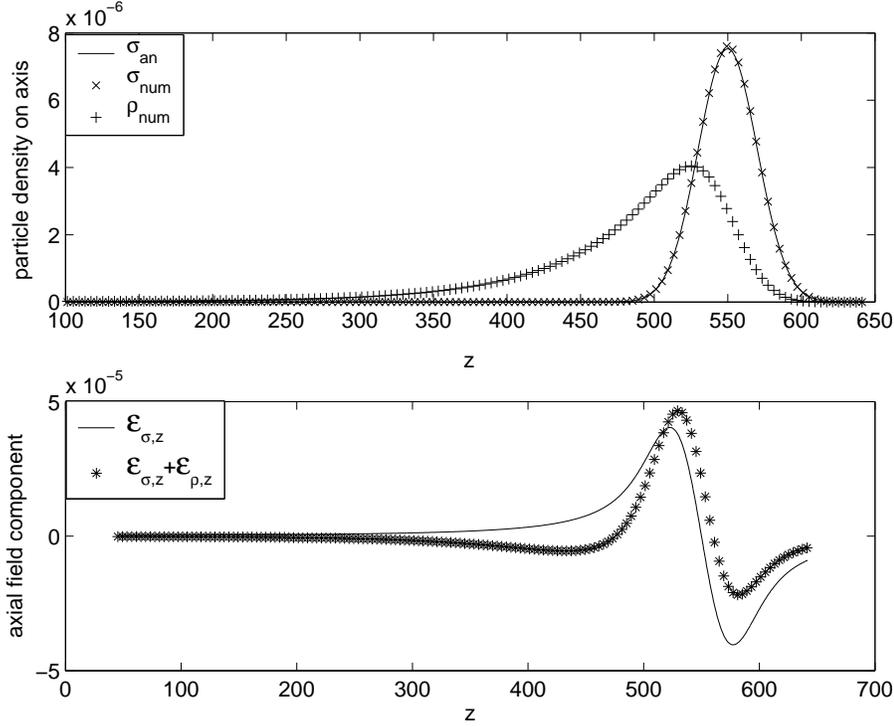}
\caption{Analytical results (solid lines) for the linearized problem compared 
to a numerical evaluation (symbols) of the full nonlinear 
model~(\ref{sigmand})-(\ref{poissnd}) in a background field 
$\E_b=-0.25\;\hat{\bf e}_z$, and $D=0.1$, $\nu=0$ and $\sigma_0=10^{-4}$. 
The time of the snapshot is $\tau=2000$.  The initial condition is 
located at ${\bf r}_0=50\;\hat{\bf e}_z$. 
Upper panel: the electron density distribution (x) and ion ${\small +}$ density 
distributions along the axis of symmetry $r$=0, as computed by a numerical 
simulation of the nonlinear model; the solid line is the analytical 
solution~(\ref{sigmaav}) of the linearized problem. 
Lower panel: the axial component of the numerically computed 
space charge field (i.e. generated by the electrons and the ions) 
$\E'=\E_\sigma+\E_\rho=\bs{\cal E}-\bs{\cal E}_{b}$ (thin line) and the analytical
result for the field $\E_\sigma$ generated by the electrons only
(thick line). The field induced by the ions is discussed in the next section and not
accounted for in $\E_\sigma$. 
We remark that this situation corresponds qualitatively to the leftmost 
column of Fig.~\ref{plotav}.}
\label{electrons}
\end{center}
\end{figure}

The axial electron density distribution for a background 
field of $\E_b=0.25$ at $\tau=2000$ 
(for N$_2$ this corresponds to a reduced electric field 
${\rm E}_b(N_0/N)=50$ kV/cm and $t$=6 ns) is illustrated 
in the upper panel of Fig.~\ref{electrons}. The analytical 
solution~(\ref{sigmaav}) of the linearized continuity 
equation~(\ref{sigmandav}) is compared to a numerical evaluation
of the full nonlinear problem~(\ref{sigmand})-(\ref{poissnd}).
The excellent correspondence between the solution of both the 
linearized and the nonlinear problem shows that, at this
time, space charge effects are negligible, so that the electrons 
still are in the avalanche phase.

\subsection{Exact result for the electron generated field ${\cal E}_\sigma$}

While density and field of the ions can only be calculated
approximately and will be treated in the next section,
the electric field $\bs{\cal E}_\sigma$ generated by the Gaussian 
electron package can be calculated exactly.

The main point is that the electron density distribution (\ref{sigmaav}) is 
spherically symmetric about the point ${\bf r}_0-\bs{\cal E}_{b}\tau$. 
The electric field ${\E}_\sigma(s,\tau)$ at the point 
\be 
{\bf s}={\bf r}-{\bf r}_0+\bs{\cal E}_{b}\tau 
\ee
can therefore be written as $\E_\sigma(s,\tau)=-\mE_\sigma(s,\tau)\hat{\bf e}_s$,
where $\hat{\bf e}_s$ is the unit vector in the radial ${\bf s}$ direction.
Its magnitude can be computed with Gauss' law of electrostatics
(in the same way as the gravitational force field of a spherically
symmetric mass distribution). It uses the fact that the field at radius $s$ is
determined by the total charge inside the sphere of radius $s$, and
independent of charges outside this radius, as long as the distribution is
spherically symmetric. It yields
\be
\hspace{-1cm}{\cal E}_\sigma(s,\tau) =  
\frac1{s^2}\int_0^{s} \sigma_0 e^{f\tau}\;
\frac{e^{-r^2/(4D\tau)}}{(4\pi D\tau)^{3/2}}\;r^2dr 
= \frac{\sigma_0 e^{f\tau}}{16\pi D\tau}\; 
F\left(\frac{s}{\sqrt{4D\tau}}\right),
\label{esigma}
\ee
with 
\begin{equation}
\label{F}
F(x)=\frac1{x^2}\;\;\frac{4}{\sqrt{\pi}}\int_0^xy^2e^{-y^2}dy
=\frac{\text{erf}~x}{x^2}-\frac{2}{\sqrt{\pi}}\;\frac{e^{-x^2}}{x},
\end{equation}
where erf is the error function. 

The spatial maximum of the field strength
${\cal E}_\sigma$ is determined by the maximum of $F(x)$; 
evaluating $F'(x)=0$ shows that it is located at an $x$ such that 
\begin{equation}
\frac{2}{\sqrt{\pi}}(x+x^3)e^{-x^2} = {\rm erf}~x.
\end{equation}
Solving this equation numerically leads to a position of the maximum
of about $x\simeq 1$ (which is the radius at which the Gaussian electron
distribution has dropped to $1/e$ of its maximal value)
and to the value $F(1)\simeq 0.4276$.  
The spatial maximum of the electron generated electric field strength becomes
\begin{equation}
{\cal E}_\sigma^{max}(\tau)
\simeq\frac{\sigma_0e^{f\tau}}{16\pi D\tau}F(1),
\label{esm}
\end{equation}
it is located on the sphere parameterized through
\begin{equation}
|{\bf r}-{\bf r}_0-\bs{\cal E}_{b}\tau|\simeq \sqrt{4D\tau}.
\label{remax}
\end{equation}

In the original cylindrically symmetric coordinate system $(r,z)$, 
the axial field component is directed in the negative
$z$-direction, i.e. in the same direction as the background field,
``ahead'' of the electron cloud ($z>z_0+\mE_b\tau$)
as is illustrated by the solid line in the lower panel of 
Fig.~\ref{electrons}. Combining this with Eq.~(\ref{remax}), we find that 
the maximal field strength $|\E_b+\E_\sigma|$ and its location are 
\be
\max_{\bf r}|\E_b+\E_\sigma|=|\E_b+\E_\sigma|({\bf r}_m,\tau)=\mE_b+\mE_\sigma^{max}(\tau) \, ,
\label{ebsmax}
\\
{\bf r}_m(\tau)\simeq (z_0+\mE_b\tau+\sqrt{4D\tau}){\bf\hat e}_z\, .
\label{rebsmax}
\ee

\subsection{A lower bound for the transition}

Since the avalanche to streamer transition takes place when 
space charge effects start to affect the electric field, we choose
to base the criterion for the transition
on the maximal relative field enhancement $k(\tau)$ defined in 
Eq.~(\ref{k}), which for the dimensionless field simply reads
\be
k(\tau)=\frac{\max_{\bf r}|\E({\bf r},\tau)|-|\E_b|}{|\E_b|}.
\label{knd}
\ee
Here $\E=\E_b+\E_\sigma+\E_\rho$
is the total electric field, $\E_\sigma$ and $\E_\rho$ being the fields of
the electrons and the ions, respectively. 
We will show in the next section that $k_t=0.03$ is
an appropriate estimate for the maximal relative field enhancement 
at the mid gap avalanche to streamer transition. At lower values of $k$,
space charge effects can be neglected, whereas at higher values 
the dynamics of the electrons are nonlinear and the full streamer 
equations~(\ref{sigmaav})-(\ref{poissnd}) have to be solved. 

As a first estimate for the space charge field, and thereby for the 
avalanche to streamer transition, we compute the 
field generated by the electrons only and neglect the ion field. This is a decent 
approximation, as the lower panel in Fig.~\ref{electrons} shows. 
Actually, the magnitude of the monopole field $\E_\sigma$ 
ahead of the electron cloud is an upper 
bound for the magnitude of the field created by the dipole of electrons on 
the one hand and the positive charges left behind by the electron cloud on 
the other hand. 
Therefore, the maximal relative field enhancement due to the electrons,
$k_\sigma(\tau)=\mE_\sigma^{max}(\tau)/\mE_b$, exceeds the transition
value after a shorter travel time $\tau_\sigma$ and distance
then the genuine relative field enhancement $k(\tau)$ of Eq.~(\ref{knd}).
Hence, $\tau_\sigma$ is a lower bound for the time 
$\tau_{a\rightarrow s}$ of the avalanche-to-streamer transition. 

The lower bound $\tau_\sigma$ for the transition can be  
expressed through Eq.~(\ref{esm}) as
\begin{equation}
f\tau_\sigma-\ln({\mE}_b\tau_\sigma)
\simeq\ln\frac{16\pi k_t D}{F(1)\sigma_0}.
\label{transt0}
\end{equation}
As travel time and travel distance are related through the drift velocity
${\mE}_{b}$,
$f(|\E_b|,\nu)\tau_\sigma$ is found to be identical to 
$(\alpha(|{\bf E}_{b}|)-\nu_a/\mu_eE_b)d_\sigma$ 
in dimensional units where $d_\sigma=\mu_eE_bt_\sigma$
is the avalanche travel distance.
In dimensional quantities, Eq.~(\ref{transt0}) takes the form
\be
\hspace{-1cm}\left(\alpha(|E_b|)-\frac{\nu_a}{\mu_eE_b}\right)d_\sigma
-\ln(d_\sigma\alpha_0)=
\ln\frac{16\pi k10^4}{F(1)}+\ln\frac{D_e\alpha_0}{\mu_eE_0}-\ln\frac{N}{N_0}.
\label{transt1dim}
\ee
For a non-attaching gas ($\nu_a=0$) at atmospheric pressure under normal
conditions with dimensionless diffusion comparable to nitrogen, 
inserting the numerical values for the parameters, we obtain
\be
\alpha(|E_b|)d_\sigma-\ln(\alpha_0d_\sigma)\approx 9.43.
\label{transt1num}
\ee
$f$ being a growing function of $|\E_b|$, Eq.~(\ref{transt0}) shows that 
the larger the field, the earlier the transition takes place, 
which is in accordance with Meek's criterion.
On the other hand, the second term on the right hand side of 
Eq.~(\ref{transt1dim}) depends on the diffusion coefficient 
in such a way that diffusion delays the
transition to streamer, as expected. 

\begin{figure}[h]
\begin{center}
\includegraphics[width=8cm]{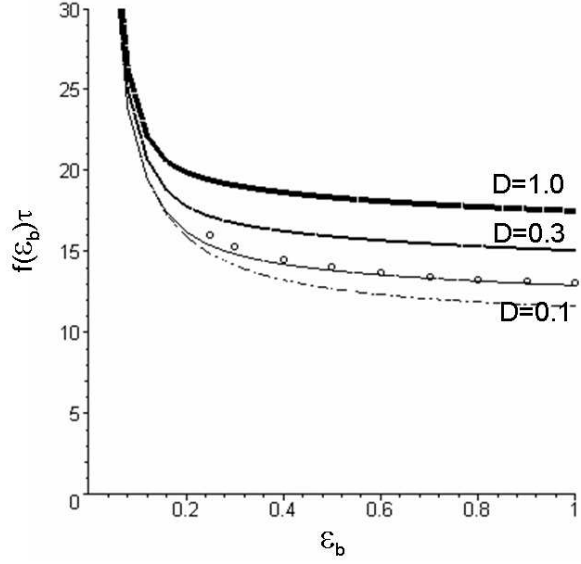}
\caption{The transition time $f\tau$ (equivalent to the travel 
distance $\alpha d$)
as a function of the background electric field
for $\sigma_0=10^{-4}$, $\nu=0$ and different values of $D$. 
Solid line: computed with Eq.~(\ref{transt1}) for $D=0.1$ (thin line), 
0.3 (medium thin line) and 1 (thickest line); 
dash-dotted line: computed with Eq.~(\ref{transt0}) for $D=0.1$; 
symbols: numerical evaluation for $D=0.1$.
Obviously, the approximation (\ref{transt1}) fits the full
numerical results very well.}
\label{transtime}
\end{center}
\end{figure}

The solution $\alpha(|E_b|)d_\sigma$ for $N_2$ at atmospheric pressure
is shown in the dash-dotted line of Fig.~\ref{transtime}, 
where it is compared to a numerical
evaluation of the transition time (circles). 
The latter have been obtained through a full simulation of the 
continuity equations~(\ref{sigmand})-(\ref{rhond}) together 
with the Poisson equation~(\ref{poissnd})~\cite{roc2002,mon2005-2}
that was also used to generate Fig.~\ref{plotav}. 
Though the qualitative features of the transition time
are well reproduced, this figure shows that the underestimation
of the transition time is significant, and that it is necessary 
to include the field of the ion trail left behind by the electrons. 

\section{Ion distribution and field}
\label{secion}
\subsection{Exact results on the spatial moments of the distributions}

To get a more accurate estimate for the avalanche-to-streamer transition, 
the field generated by the positive and negative ions has to be
included. In the case of the ion distribution, closed analytical
results cannot be found, in contrast to the electron distribution
(\ref{sigmaav}). However, arbitrary spatial moments of the distribution
\be
\langle {\cal O}\rangle_\rho
=\frac{\int {\cal O}\;\rho\; d^3{\bf r}}{\int\rho \;d^3{\bf r}},
~~~{\rm where }~~{\cal O}=z^n~~{\rm or}~~r^n,
\label{defmom}
\ee
can be derived analytically. Here $z$ is the direction of the homogeneous 
field ${\cal E}_{b}$ and $r$ is the radial direction.
First, the evolution equation (\ref{rhond}) 
for the ion density $\rho$ is integrated in time and the analytical form 
(\ref{sigmaav}) for $\sigma({\bf r},\tau)$ is inserted. 
As $f=f(|\E_{b}|,\nu)$ is constant in space and time  
one finds
\be
\label{rho}
&&\rho({\bf r},\tau)-\rho({\bf r},0)=
\\
&&f \sigma_0 \int_0^\tau d\tau'\;e^{f\tau'}\;
\frac{e^{-(z-z_0 -{\cal E}_{b}\tau')^2/(4D\tau')}}
{\sqrt{4\pi D\tau'}}\;
\frac{e^{-r^2/(4D\tau')}}{4\pi D\tau'}.
\nn
\ee
Here the initial perturbation is located at $z_0$ on the axis $r=0$.
The moments (\ref{defmom}) can now be derived from (\ref{rho})
by exchanging the order of spatial and temporal integration.
In particular, one finds
\be
\int\rho \;d^3{\bf r}&=&\sigma_0\;e^{f\tau},\\
\int z\;\rho \;d^3{\bf r}&=&\sigma_0\;e^{f\tau}\;
\left(z_0+{\cal E}_{b}\tau-\frac{1-e^{-f\tau}}{f/{\cal E}_{b}}\right),
\nn
\ee
and higher moments can be calculated in the same way.
For the moments of $\rho$ in the axial direction, this gives
\be
\langle z\rangle_\rho&=&
z_0+{\cal E}_{b}\left(\tau-\frac1{f}\right) +O\left(e^{-f\tau}\right),
\label{z1rho}
\\
\langle z^2\rangle_\rho - \langle z\rangle_\rho^2
&=&\left(\frac{{\cal E}_{b}}{f}\right)^2+2D\left(\tau-\frac1{f}\right)
+O\left(e^{-f\tau}\right).
\ee
The second moment of $\rho$ in the radial direction is
\be
\langle r^2\rangle_\rho = 2D\left(\tau-\frac1{f}\right)
+O\left(e^{-f\tau}\right).
\ee

For comparison, the moments of the Gaussian electron distribution 
(\ref{sigmaav}) are easily found to be
\be
\label{e1}
\langle z\rangle_\sigma&=& z_0+{\cal E}_{b}\tau,\\
\langle z^2\rangle_\sigma - \langle z\rangle_\sigma^2
&=&2D\tau,\\
\langle r^2\rangle_\sigma &=& 2D\tau.
\label{e3}
\ee

\subsection{Discussion of the moments}

Let us now interprete these moments. A first moment of a spatial 
distribution gives its center of mass.
For the second moment, the cumulant
\be
\langle z^2\rangle^c_x:=
\left\langle \big(z - \langle z\rangle_x\big)^2\right\rangle_x
=  \langle z^2\rangle_x - \langle z\rangle_x^2,
~~~x=\sigma, \rho.
\ee
measures the quadratic extension from the center of mass.
As the center of mass lies on the axis, for the radial
extension the distinction between
second moment and its cumulant need not be made.

The moments for the electrons (\ref{e1})--(\ref{e3}) 
have a simple structure: the center of mass of the electron
package is located at $z=z_0+{\cal E}_{b}\tau$, and the package
has a diffusive width $\sqrt{2D\tau}$ around it, 
both in the forward $z$ direction  and in the radial $r$ direction.

The ion cloud shows a more complex behavior; it is evaluated close
to the avalanche-to-streamer transition where $f\tau=\alpha d=O(10)$,
therefore the terms of order $e^{-f\tau}$ are neglected.

First it is remarkable that the center of mass of the ion cloud~(\ref{z1rho}) 
shifts with precisely the same velocity as the electron cloud 
though the ion motion is neglected while the electrons drift, 
therefore the ion center of mass 
is at an approximately constant distance ${\cal E}_{b}/f$ 
behind the electron center of mass. This distance 
\be
\ell_\alpha=\frac{{\cal E}_{b}}{f({\cal E}_{b})}=\frac{\alpha_0}{\alpha(E_{b})}
\ee
corresponds to the dimensional ionization length $1/\alpha(E_{b})$.

The quadratic radial width of the ion cloud $2D(\tau-1/f)$
is $2D/f$ smaller than the one of the electron cloud. This is related
to the fact that the electron cloud also was more narrow at the earlier
times when it left the ions behind. The ion cloud is more extended in the
$z$ direction. More precisely, its length is $\ell_\alpha$ larger
than its width. This comes from the ions being immobile, 
therefore a trace of ions is
left behind by the electron cloud. Moreover, it can be remarked that 
the difference between quadratic width and length of the ion cloud is 
given by the same ionization length $\ell_\alpha$ as the distance 
between the centers of mass of the ion and the electron cloud. 
We refer to the left column of  Fig.~\ref{plotav} 
for the illustration of these density distributions.

\subsection{An estimate for the transition}

One can assume as in \cite{baz1998} that the ions have a spatial distribution 
similar to the electrons, thus a Gaussian with the same width as 
the electron cloud, but centered around $(r=0,z=\langle z\rangle_\rho)$: 
\be
\rho_1(r,z,\tau)=\sigma_0\,e^{f\tau}\;
\frac{e^{-\big[(z-\langle z \rangle_\rho)^2+r^2\big]/(4D\tau)}}
{(4\pi D\tau)^{3/2}}.
\label{rho1}
\ee
In this approximation, the total electric field becomes:
\be
\hspace{-1cm}\bs{\cal E}_1(r,z,\tau) = \bs{\cal E}_{b} 
-\frac{\sigma_0 e^{ft}}{16\pi \;D\tau}
    \left[F\left(\frac{|{\bf s}_\sigma|}{\sqrt{4D\tau}}\right)
     \frac{\bf s_\sigma}{|{\bf s_\sigma}|}
        + F\left(\frac{|{\bf s}_\rho|}{\sqrt{4D\tau}}\right)
    \frac{\bf s_\rho}{|{\bf s_\rho}|}
    \right],
\label{E1}
\ee
where
\ba
{\bf s}_x & = & {\bf r} - \langle z\rangle_x\;\hat{\bf e}_z
~~~{\rm for}~~x=\rho,\sigma\label{rs}\label{rr}
\ea
are the distances to the electron and ion 
centers of mass.

The maximum of the field ${\cal E}_1$ can not be computed analytically. 
However, in Fig.~\ref{plotav} and in the lower panel of Fig.~\ref{electrons},
it can be seen that this maximum is located on the axis ahead
of the electron cloud, and that  
the location of the maximum of the total field and that of the electron field nearly 
coincide. This can easily be explained physically: the total field is the 
sum of the fields induced by the electrons and by the ions. Its maximum is located
just ahead of the electron cloud, where the electron field
varies rapidly, while the field contribution of the ions is smoother 
and smaller as we are interested in its contribution further away 
from the center of the ion distribution.
Therefore the maximum position of the total field is essentially
identical to the maximum position of the electron field.
This justifies our approximation to evaluate the field 
${\cal E}_1$ at the maximum position ${\bf r}_m$ of $|\E_b+\E_\sigma|$ as defined 
in Eq.~(\ref{rebsmax}).
The maximum of the electric field can thus be approximated as:
\ba
&&{\cal E}_1^{max}(\tau) 
\simeq {\cal E}_1(r=0,z=z_0+\E_b\tau+\sqrt{4D\tau},\tau) \nonumber \\
&&~~~   =  {\cal E}_{b}+\frac{\sigma_0e^{f\tau}}{16\pi D\tau}
    \left[F(1)-F\left(1+\sqrt{\frac{\ell_\alpha^2}{4D\tau}}\right)\right].
\label{e1max}
\ea
Then ${\cal E}_1^{max}-{\cal E}_{b}=k{\cal E}_{b}$ implies for the transition time $\tau_1$:
\be
f\tau_1-\ln({\cal E}_b\tau_1)
-\ln\frac{F(1)}{F(1) - F\left(1+ \sqrt{\frac{\ell_\alpha^2}{4D\tau_1}}\right)}
=\ln\frac{16\pi kD}{F(1)\sigma_0},
\label{transt1}
\ee
where $F(x)$ is defined in Eq.~(\ref{F}).
The argument of the logarithm in the third term on the left hand side is 
larger than 1, therefore this criterion gives a later transition time than that
based on the field of the electrons only. This is what we expect considering
that the ions tend to reduce the field of the electrons, thus the effect
of space charge. The correction given by the ion field is a function
of the ratio of the ionization length $\ell_\alpha$ and the diffusion 
length $\sqrt{2D\tau}$. At early times, this ratio goes to infinity, 
and the correction given by the ion cloud is negligible. However, at
later times, the correction becomes more significant.

\subsection{The analytically approximated transition criterion
compared with numerical results}

We now compare again our analytical results for the linearized problem
to the outcome of numerical simulations of the full nonlinear 
model~(\ref{sigmand})-(\ref{poissnd}). 

\begin{figure}[t]
\begin{center}
\hspace{-0.5cm}\includegraphics[width=9cm]{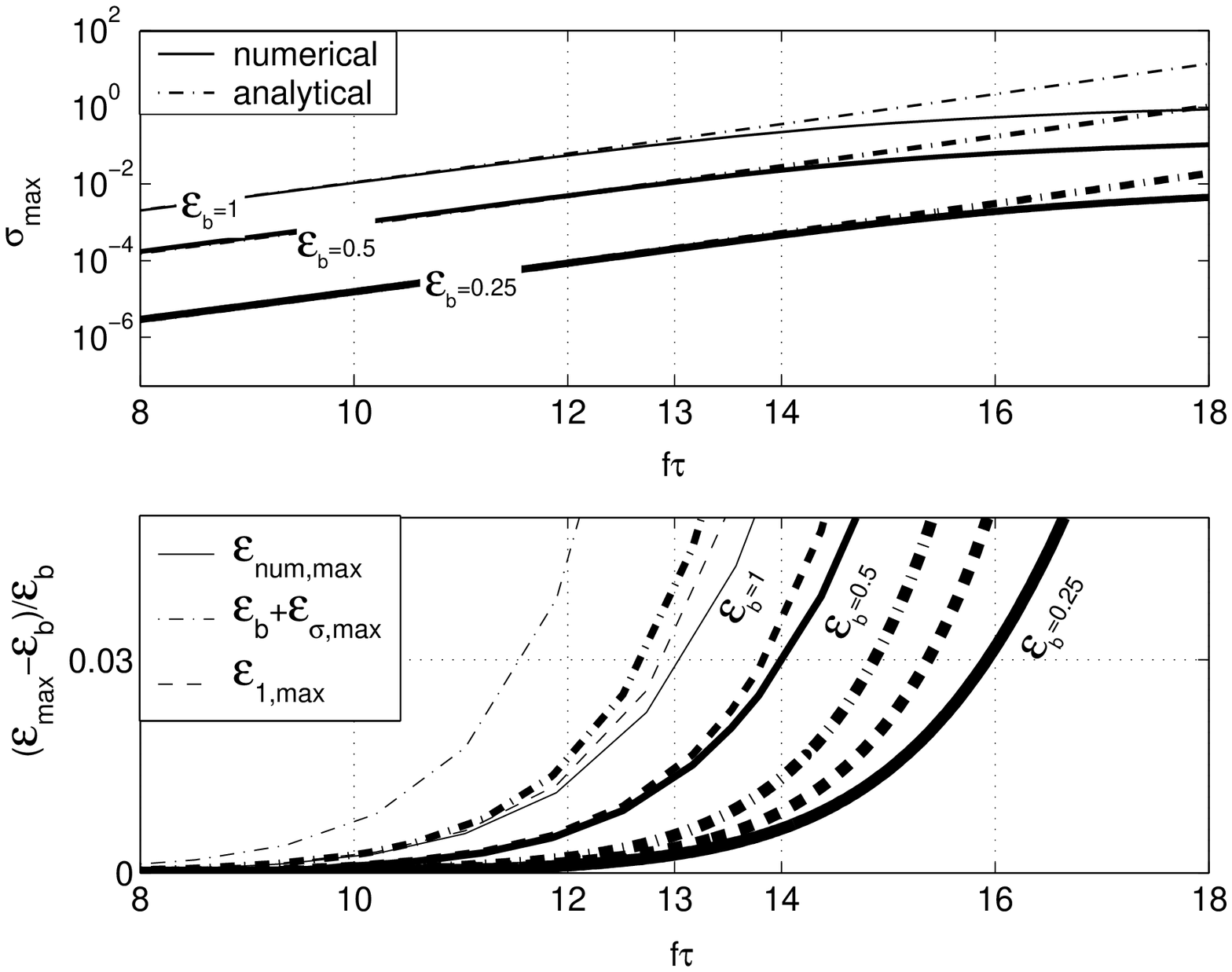}
\caption{Comparison of the analytical approximations and simulation results 
of the full non linear streamer equations~(\ref{sigmand})-(\ref{poissnd}), 
for various background electric field strengths $\mE_{b}$ ; thin line: $\mE_b=1$, 
medium thick line $\mE_b=0.5$ and thick line $\mE_b=0.25$. \\
Upper panel: the evolution of the maximal electron density as a function of $f(\mE_b\tau)$
as computed within the full nonlinear 2D model (solid lines) and as
given by the analytical solution~(\ref{sigmamaxth}) of the linearized problem 
(dash dotted lines).\\
Lower panel: the evolution of the maximal electric field enhancement 
$k=(\mE_{max}-\mE_b)/\mE_b$ as a function of $f(\mE_b\tau)$. Solid 
lines: numerical solution of the full nonlinear 2D model; dashed dotted lines:
only the field of the electrons is accounted for in the analysis, see Eq.~(\ref{esm});
dashed lines: analytical approximation~(\ref{e1max}) of the total field.}
\label{comperel}
\end{center}
\end{figure}

In the upper panel of Fig~\ref{comperel} the evolution of the maximal electron density 
as a function of $f(|\E_b|)\tau$ is shown. 
Numerical and analytical solutions coincide
during the avalanche phase, but deviate eventually. This enables us to estimate the 
moment at which the space charge effects set in, and thus when the streamer regime is reached.
In the lower panel of Fig~\ref{comperel} the evolution of the maximal relative field enhancement
is considered.  Looking at the simulation results (the solid lines), we see that $k=0.03$ 
gives a good estimate of the transition time.

The approximation~(\ref{e1max}) for the maximal field has now become 
much better than the previous approximation~(\ref{esm}) based on the electron cloud
only. Indeed, for e.g
the case of $\mE_b=0.5$ (corresponding to the medium thick lines), 
the numerically 
computed field (solid line) reaches the transition value ($(\E_{num}-\E_b)=0.03\E_b$ at 
$f\tau\approx14$. When only the field of the electrons is taken into account, this
value would already be reached at $f\tau\approx12.6$, while the correction based
on the approximation of the ion cloud leads to a transition time of $f\tau\approx13.9$. 
The correction becomes especially important at higher fields. In low fields,
the approximation of the ions shows somewhat larger deviations. 
We notice that the analytical approximation 
$\rho_1$ is narrower and higher than the genuine one, and therefore leads to an 
overestimation of the field generated by the ions inside the ion cloud.  For an even more 
accurate estimate of the total field between the electron and the ion cloud
we refer to Appendix 2, where it is also shown this will not lead to a significant 
improve in the estimate of the maximal field ahead of the electron cloud.

In Fig.~\ref{transtime} we compare the transition times given by Eqs.~(\ref{transt0}) and
(\ref{transt1}) with numerically evaluated transition times. It shows that the
approximation of similar electron and ion distributions leads to a very good 
approximation of the transition time. This figure also illustrates that 
the transition time $f\tau$ depends strongly on the electric field, and increases
for smaller fields. 
Moreover, looking at the transition time for higher diffusion coefficients, 
it is seen that  
diffusion tends to delay the transition to the streamer regime. This can be
expected, since diffusion will tend to broaden the electron cloud, thereby
suppressing space charge effects. Depending on the external parameters,
the value of $\alpha d$ at the time the transition takes place can vary 
by a factor two or more.

\subsection{The final results on the transition criterion}

\begin{figure}[h]
\begin{center}
\includegraphics[width=9cm]{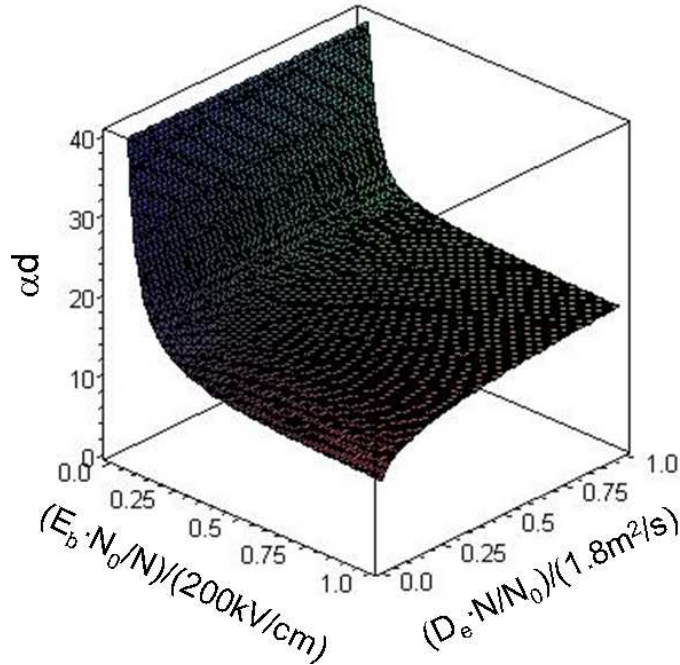}
\caption{The transition distance $\alpha d$ of an electron avalanche
in a non-attaching gas ($\nu=0$) like $N_2$ or $Ar$ according to Eq.~(\ref{transt1}) as a function 
of the background electric field ${\cal E}_{b}$ 
and the diffusion coefficient $D$ for $\sigma_0=10^{-4}$. The values largely deviate
from the Raether-Meek criterion~(\ref{meek}).}
\label{transED}
\end{center}
\end{figure}

The transition time approximated by Eq.~(\ref{transt1}) as a function 
of both background electric field and diffusion coefficient is visualized 
as the 3-dimensional in Fig.~\ref{transED}. This figure shows that
the Raether-Meek transition criterion, that stated that $f\tau$ 
takes an approximately constant value of 18 to 21,
corresponds to the case of relatively high diffusion and background field. 
However, realistic values of $D$ are smaller than unity, 
and a background electric field higher 
than 2 also leads to unrealistic values. So in the parameter range 
of real experiments, the correction given by Eq.~(\ref{transtime}) 
on the Raether-Meek criterion can not be neglected.

\begin{figure}
\begin{center}
\includegraphics[width=7cm]{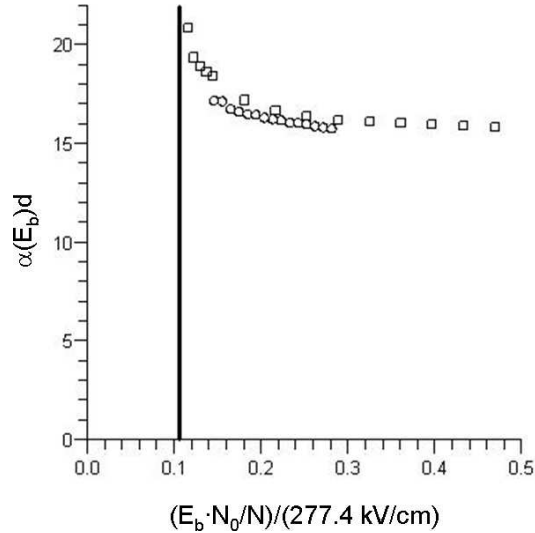}
\caption{The value of $\alpha d$ when the electron avalanche has
traveled a distance $d$ and has reached the point at which the transition 
to streamer takes place. The x-axis shows the dimensionless electric field.
The line at $\E\simeq 0.01$ indicates the field below which attachment overcomes 
ionization, and hence there is no electron multiplication. $\circ$: experimental value
for $\alpha(E)$~\cite{dut1975} and $D/mu_e$~\cite{lak1977}; $\square$: values from the Boltzmann 
solver~\cite{liu2004}.}
\label{ftair}
\end{center}
\end{figure}

We no discuss the particular example of an electron avalanche in (dry) air, 
for which different coefficients have to be used than in N$_2$. In particular, the ionization length and field in air are given by~\cite{rai1991}
$\alpha_0=0.87 \mu{\rm m}(N/N_0)$ and $E_0=277.4{\rm kV/cm}\cdot N/N_0$. 
For the values of the mobility and the diffusion coefficient of the electrons
as a function of the electric field we use experimental values as well as 
numerical solutions of the Boltzmann equation (see Appendix A). 
Inserting those in Eq.~(\ref{transt1}), we can compute the value of $\alpha(|{\bf E}|)d$ 
at the transition for different background fields, showed in  Fig.~\ref{ftair}.
At large fields, the value of $\alpha d$ at transition saturates towards 16, and 
grows asymptotically as the reduced field approaches a value of $\pm$27.7 kV/cm. At even 
lower fields attachment overcomes electron impact ionization,
and a single electron can not generate a streamer.
Large values of $\alpha d$ as in Fig~\ref{transED} are not found,
as electron attachment limits the emergence of streamers in low fields
(see Eq.~(\ref{Ek})). So for air, $\alpha d$ drops from 21 to 15 with
growing field.  
 
\section{Summary and conclusions}

Recent simulations (see Fig.~\ref{plotav}) have shown that 
an electron avalanche turns into a streamer when
the field enhancement due to space charges 
is about 3\%. In this paper, the theory behind the commonly
used Raether-Meek criterion is reviewed, as it assumes
a linear behavior of the electrons (i.e. space charge effects can be neglected),
which is in contradiction with the requirement 
of the space charge field to be in the same order of magnitude as 
the background field for the transition to occur.
    
A dimensional analysis has been carried, enabling us to 
identify the characteristic length scales, which are a function of the 
neutral gas density. In particular, the dimensionless quantities  
$\alpha(\mE_b)d$ and $D=D_e\alpha_0/(\mu_eE_0)$ have been extracted from
the problem.  The first gives the distance $d$ in multiples
of the effective ionization length in the background field, while the 
latter gives the ratio between diffusive and advective
transport of electrons. The continuity equations for the positive 
and negative ions~(\ref{n+d})-(\ref{n-d})
reduce to one single equation~(\ref{rhond}) holding both positive 
and negative ions after rescaling, making
the further analysis valid for both attaching gases like air or non-attaching
gases like N$_2$ or Ar. 

The avalanche regime was identified as the phase during which
space charge effects are negligible. This implies that the problem
can be linearized around the background field, making it 
well-suited for analytical treatment. Indeed, for an electron avalanche 
evolving in a homogeneous background electric field, a closed analytical
expressions exist for the density distribution of the electrons.
Comparing this analytical solution of the linearized problem to the results of
a numerical simulation of the full nonlinear problem, it could
be concluded that the transition to streamer takes when the 
maximal relative field enhancement $k$ has reached a value of approximately 3\%.

We have shown that the electric field of the electron cloud 
during the avalanche regime can also be described by a closed
expression. This led to the derivation of a lower bound for the 
avalanche to streamer transition~(\ref{transt0}). 
The estimate of the transition time has been improved by taking 
into account the field of the ions for which, in contrast to the electrons, 
no closed expression exists. However, the contribution 
of the ions to the maximal relative field enhancement can be well approximated.
leading to an analytical estimate of the avalanche to streamer transition~(\ref{transt1}).

The transition distance $\alpha d$ strongly depends on diffusion $D$ and on
the background electric field. For high fields, the transition time saturates
towards $\alpha d\simeq15$. On the other hand, for low fields,
when the processes are diffusion dominated, the avalanche lasts longer.
We remark that the striations observed in~\cite{dow2003} 
are generic for atomic gasses with essentially only elastic and
ionizing collisions, i.e. with very few inelastic processes~\cite{bro2005}.

In air, attachment limits the emergence of a streamer in low fields (see Eq.~(\ref{Ek})). 
In this case, $\alpha d$ at transition is in the range of 16 (for high background fields) 
to 21 (for fields approaching $E_k$). It is remarkable that in the
end, due to attachment cut-offs, Meek's criterion performs quite good for 
the emergence of streamers in free space. 
In non-attaching gases like N$_2$ or Ar, the correction on Meek's criterion,
that stated that $\alpha d\simeq 18~-~21$, becomes important at low fields.
There the relatively strong diffusion delays the transition to streamer
considerably. We emphasize that the use of dimensionless quantities enable 
us to translate the criterion given in~(\ref{transt1}) to any given neutral gas type
and density. Evaluating the characteristic scales for these conditions,
dimensionless field and diffusion can be computed, and the value of $\alpha d$ at
transition can be computed from Eq.~(\ref{transt1}) or read from Fig.~\ref{transED}.
Actually, Fig.~\ref{transED} can also be used for attaching gases, as long as  
the ionization threshold field $E_k$ is accounted for.

The analytical models presented in this paper give a useful tool to describe the 
streamer formation. We stress that our criterion for the transition is based on 
a significant contribution of the space charges on the background electric field.
Our analysis fully relied on the linearization of the streamer equations around 
the background field. The nonlinear streamer propagation is the subject of other 
studies. In that phase the space charges and electric field strongly
interact, and the analytical study of such streamers~\cite{meu2004} is far more difficult
than the analysis of the linear avalanche phase.
\ack{
C.M.\ acknowledges a Ph.D.\ grant of the Dutch NWO/FOM-program
on Computational Science.}
\\

\appendix
\section{Mobility and diffusion coefficients of electrons in air}

\begin{figure}[h]
\begin{center}
\includegraphics[width=6cm]{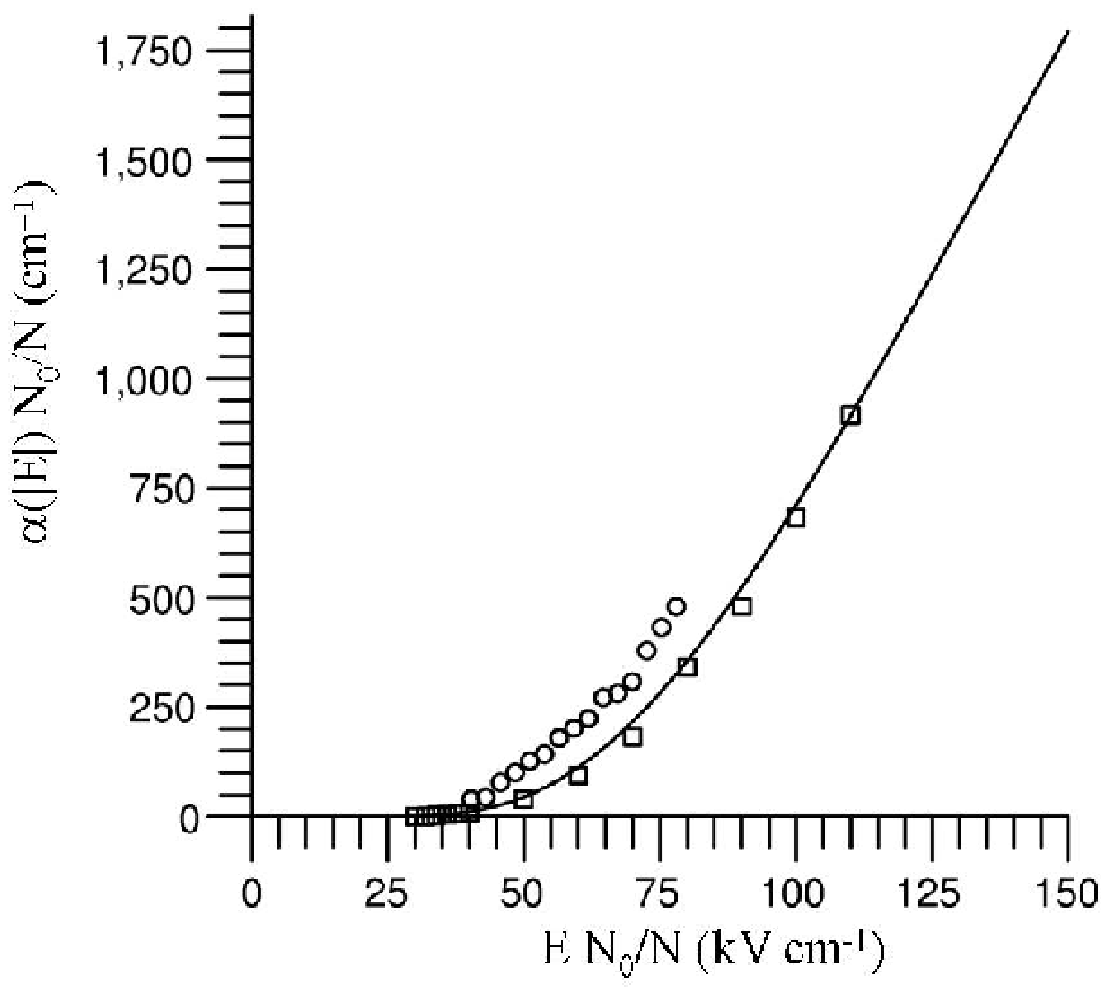}
\includegraphics[width=6cm]{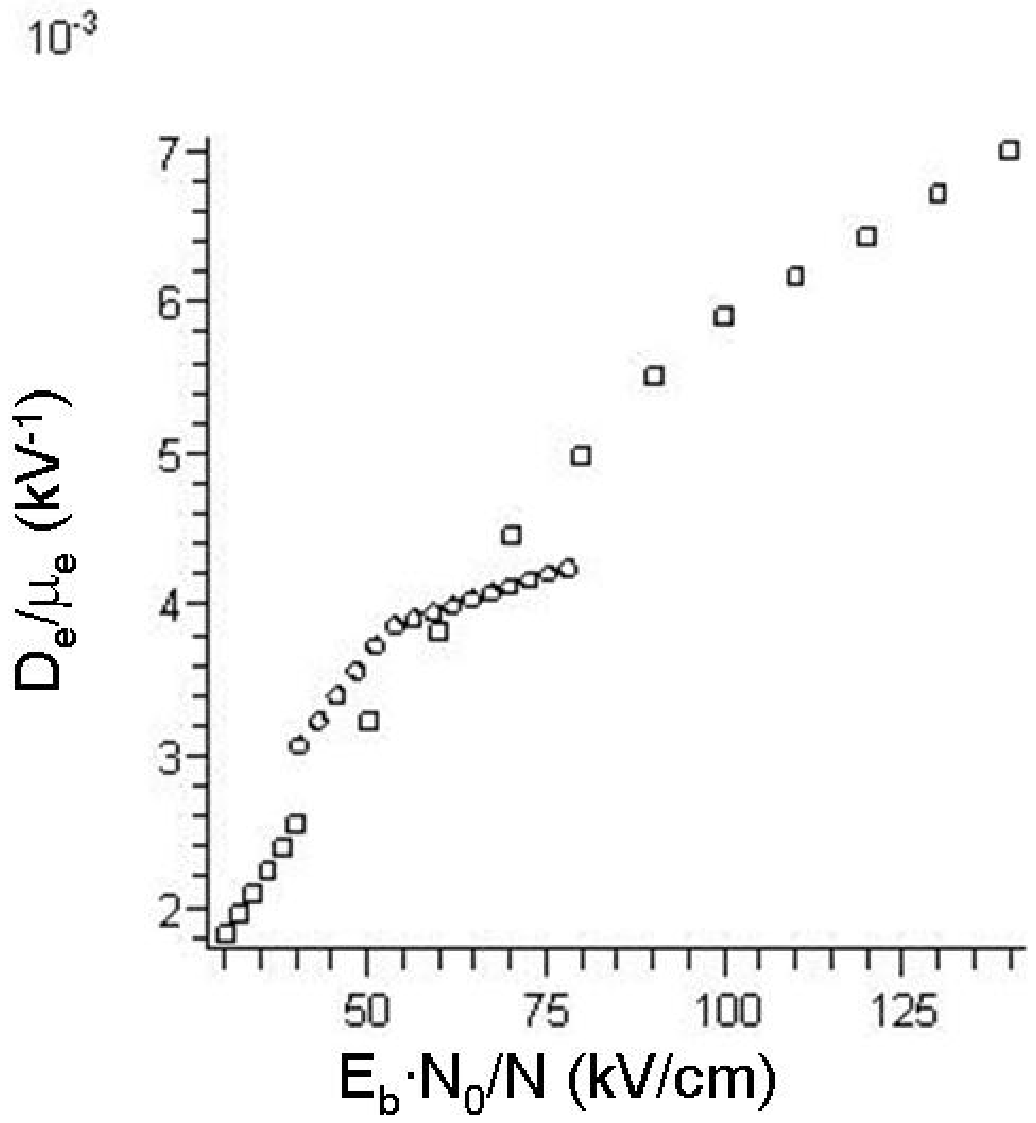}
\caption{The ionization coefficient (left) and ration of electron diffusion and mobility (right)
in air, as a function of the reduced electric field. $\circ$: experimental measurements 
(the values for $\alpha$ are taken from~\cite{dut1975}, the values for $D/\mu_e$ from~\cite{lak1977}.
$\square$: solution of the Boltzmann equation~\cite{liu2004}. The solid line shows the ionization
coefficient following the empirical formula~(\ref{alpha}) given in~\cite{rai1991},
with $\alpha_0=0.87 \mu{\rm m}(N/N_0)$ and $E_0=277.4{\rm kV/cm}\cdot N/N_0$.}
\label{coeffair}
\end{center}
\end{figure}

To compute the transition time in air, we use values of the electron mobility 
and diffusion coefficient found in literature. 
In the left plot
of Fig.~\ref{coeffair} measured and calculated values of $\alpha(|E|)$ are given,
as well as the fit $\alpha(|E|)=\alpha_0\exp(-E_0/|E|)$ are shown. 
The experimental values have been found in the survey of electron swarm data by Dutton~\cite{dut1975}.
The computed values are the solution of the Boltzmann equation and have been taken from~\cite{liu2004}.
Also, the empirical approximation of the ionization coefficient as a function of the
background field as given by~\cite{rai1991} is shown, $\alpha(|{\bf E}|)=\alpha_0\exp(-E_0/|E|)$
with $\alpha_0=0.87 \mu{\rm m}(N/N_0)$ and $E_0=277.4{\rm kV/cm}\cdot N/N_0$.

The values of $D_e/\mu_e$ as a function of the reduced electric field are given in
the right plot of  Fig.~\ref{coeffair}. Again, computed values from~\cite{liu2004}
are shown, as well as measured values found in~\cite{lak1977}. The value
of the diffusion coefficient as a function of the electric field can obviously easily 
be derived from these figures.

\section{A more accurate approximation for the ion density distribution}

The approximation for the ion distribution $\rho_1$ in Sect.~\ref{secion} leads to a relatively good
approximation for the transition time in the case of a mid-gap transition. However,
the real spatial distribution of ions is more narrow in the $r$-direction, and can be wider
and asymmetrical in the $z$-direction.  In this appendix we present another
approximation for the ion distribution, which will lead to a better overall approximation
of the electric field, and of the self field induced by the ion trail. The price however to pay for this is a much more
complicated analytical expression for the density and the field.

A better approximation for $\rho$ would then be an ellipsoidal Gaussian
distribution centered around $(r=0,z=\langle z\rangle_\rho)$ with width 
$\langle z^2\rangle_\rho^c=\langle z^2\rangle_\rho-\langle z\rangle_\rho^2$ and 
$\langle r^2\rangle_\rho^c=\langle r^2\rangle_\rho$ in the
$z$- and $r$-direction, respectively. The height of this Gaussian should be such that
the total amount of ions at time $t$ is still equal to $\sigma_0e^{ft}$. The appropriate
expression for the ion distribution is:
\be
\rho(r,z,t)=\frac{\sigma_0e^{ft}}{(2\pi)^{3/2}S_r^2S_z}
            e^{-r^2/(2S_r^2)-(z-\langle z\rangle_\rho)^2/(2S_z^2)}
\label{rhoell}
\ee
However, as far as we know, no closed analytical expression is known for the field 
of such an ellipsoidal Gaussian charge distribution. So instead, we take a spherical
Gaussian distribution with the same height as the one defined in Eq.~(\ref{rhoell}):
\be
\rho_2(r,z,\tau)=\frac{\sigma_0e^{f\tau}}{(2\pi)^{3/2}S_\rho^3}
            e^{-(r^2+(z-\langle z\rangle_\rho)^2)/(2S_\rho^2)}\, ,
\label{rho2}
\ee
where
\ba
S_\rho^3 & = &\langle r^2\rangle_\rho^c\sqrt{\langle z^2\rangle_\rho^c} \nonumber\\
       & = & \left(2D(\tau-\frac{1}{f})\sqrt{2D(\tau-\frac{1}{f})+l_\alpha^2}\right)\, .
\ea

The electric field induced by this ion distribution is:
\be
E_{\rho_2}(r,z,\tau)=\frac{\sigma_0e^{f\tau}}{8\pi S_\rho^2}F\left(\sqrt{\frac{|{\bf s}_\rho|^2}{2S_\rho^2}}\right)\, ,
\label{erho2}
\ee
where ${\bf s}_\rho$ is defined in Eq.~(\ref{rr}). 

\begin{figure}
\begin{center}
\includegraphics[width=9cm]{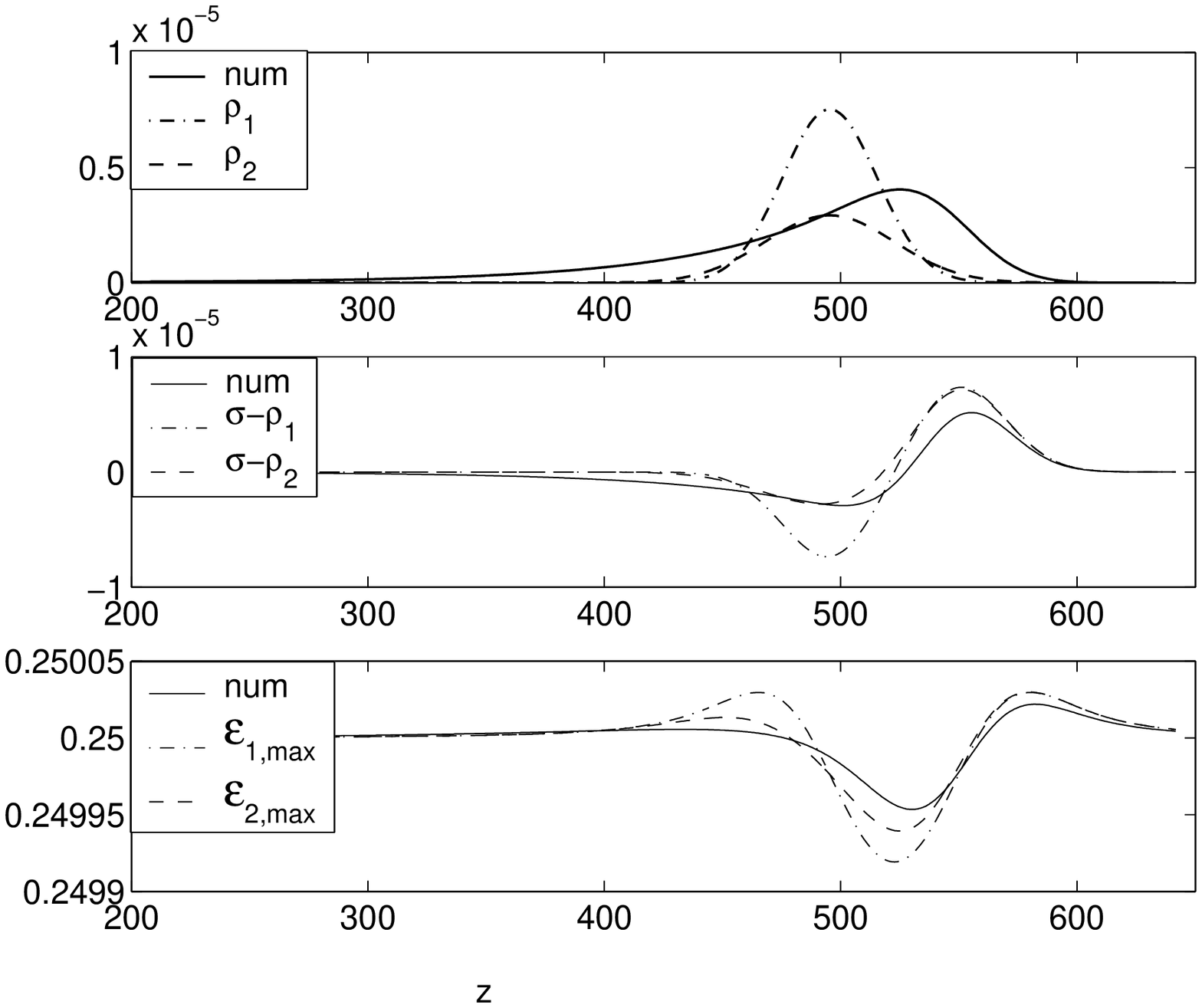}
\caption{The ion density (upper figure), total charge density (middle figure) and electric 
field (lower figure) on the axis, computed with $E_0=0.25$, at $\tau=2000$. The solid
lines give the numerical solution, the dash-dotted lines the solution 
corresponding to $\rho_1$ and the dotted lines to $\rho_2$.}
\label{compeas}
\end{center}
\end{figure}

In Fig.~\ref{compeas} we compare the densities and fields given by the numerical
solution and $\rho_1$ and $\rho_2$. It shows clearly that the approximation $\rho_2$
does not give a better approximation of the field ahead of the electron cloud. This can 
be explained by the fact that, the region ahead of the electron cloud does not contain any
ions, so that the field induced by the ions is only a function of the total number
of ions, which is the same in both $\rho_1$ and $\rho_2$. On the other hand, inside
the ion cloud the approximation is much better. Therefore, evaluating the electron
and ion densities with Eqs.~(\ref{sigmaav}) and (\ref{rho2}) and their fields
with Eqs.~(\ref{esigma}) and (\ref{erho2}), at the transition time $T_1$ given by
Eq.~(\ref{transt1}), will give a good approximation of the status of the process at the
time that streamer regime is entered.

\section*{References}

\end{document}